\documentclass[aps,a4paper,preprint]{revtex4}%
\usepackage{amsfonts}
\usepackage{amsmath}
\usepackage{amssymb}
\usepackage[]{graphicx}%
\providecommand{\U}[1]{\protect\rule{.1in}{.1in}}
\begin{document}
\title{Rising obstacle in a one-layer granular bed induced by continuous vibrations:
two dynamical regimes governed by vibration velocity}
\author{Hui Zee Then, Teruyo Sekiguchi, and Ko Okumura}
\affiliation{Physics Department and Soft Matter Center, Ochanomizu University, Japan}
\date{\today}

\begin{abstract}
Rising motion of an obstacle in a vibrated granular medium is a classic
problem of granular segregation, and called the Brazil nut (BN) effect. The
controlling vibration parameters of the effect has been a long-standing
problem. A simple possibility that the BN effect can be characterized solely
by vibration velocity has recently been pointed out. The issue has become
controversial before a long history of research, with only a few systems have
provided for the simple possibility. Here, we investigate the rising motion of
an obstacle in a vertically positioned one-layer granular bed under continuous
vibrations. We find the rising motion is composed of two distinct regimes, and
the first and second regimes are both governed, in terms of vibration
parameters, solely by the vibration velocity. We further demonstrate simple
scaling laws well describe the two regimes. Our results support the emergent
possibility on the controlling parameters of the BN effect and suggests that
this feature would be universal. We propose two possible mechanisms of
convection and arch effect for the two distinct regimes and demonstrate these
mechanism explain the scaling laws followed by our experimental data.

\end{abstract}
\maketitle

\section{Introduction}

When a cell that contains mixture of grains of large and small sizes are
vibrated, large ones tend to rise up in the cell. This effect is a typical
example of size segregation of grains by shaking, known as the Brazil nut (BN)
effect \cite{harwood1977powder,5williams1976segregation}. Since pioneering
studies in simulation \cite{6rosato1987brazil} and experiments
\cite{clement1992experimental,1KnightJaegerNagel1993,duran1993arching}, many
studies have been performed in the field of physics. As a result a number of
physical mechanisms of the phenomenon have been proposed, which include void
filling \cite{7jullien1992three}, convection \cite{1KnightJaegerNagel1993},
and arching effect \cite{duran1993arching}.

However, as for the controlling vibration parameters of the BN effects, our
understanding has become controversial. It has been believed that the
characterization of the convection-driven rising motion requires at least two
vibration parameters, the acceleration and frequency
\cite{knight1996experimental,vanel1997rise}. However, several years ago, it
was clearly shown that the convection-driven rising motion is characterized,
in terms of vibration parameters, solely by the vibration velocity, in a wide
range of experimental parameters \cite{4hejmady2012scaling}. Subsequently, a
detailed study on the granular convection induced by vibration revealed that
the convection velocity is well characterized by the velocity
\cite{yamada2014scaling}, supporting the recent study
\cite{4hejmady2012scaling}. However, despite a long history of research, the
vibration-velocity governed BN effect has been reported only in a few cases
\cite{4hejmady2012scaling,umehara2020rising}, and the problem of controlling
vibration parameters has become controversial (e.g., well-known MRI studies
\cite{ehrichs1995granular,knight1996experimental} did not demonstrate their
data can be characterized solely by vibration velocity). Therefore, providing
experimental data that can be characterized solely by vibration velocity in
different systems is an emergent important issue for settling this
long-standing issue.

In order to elucidate the controlling vibration parameters in the BN effect,
we investigate the rising motion in a one-layer granular bed under continuous
vibration. As a result, we find that the dynamics are divided in first and
second regimes, and both are well characterized solely by the vibration
velocity through simple scaling laws. In addition, we propose possible
mechanisms of convection and arch effect for the two regimes, and demonstrate
these mechanisms explain the scaling laws followed by the experimental data.

\section{Experiment}

\begin{figure}[ptbh]
\includegraphics[width=\textwidth]{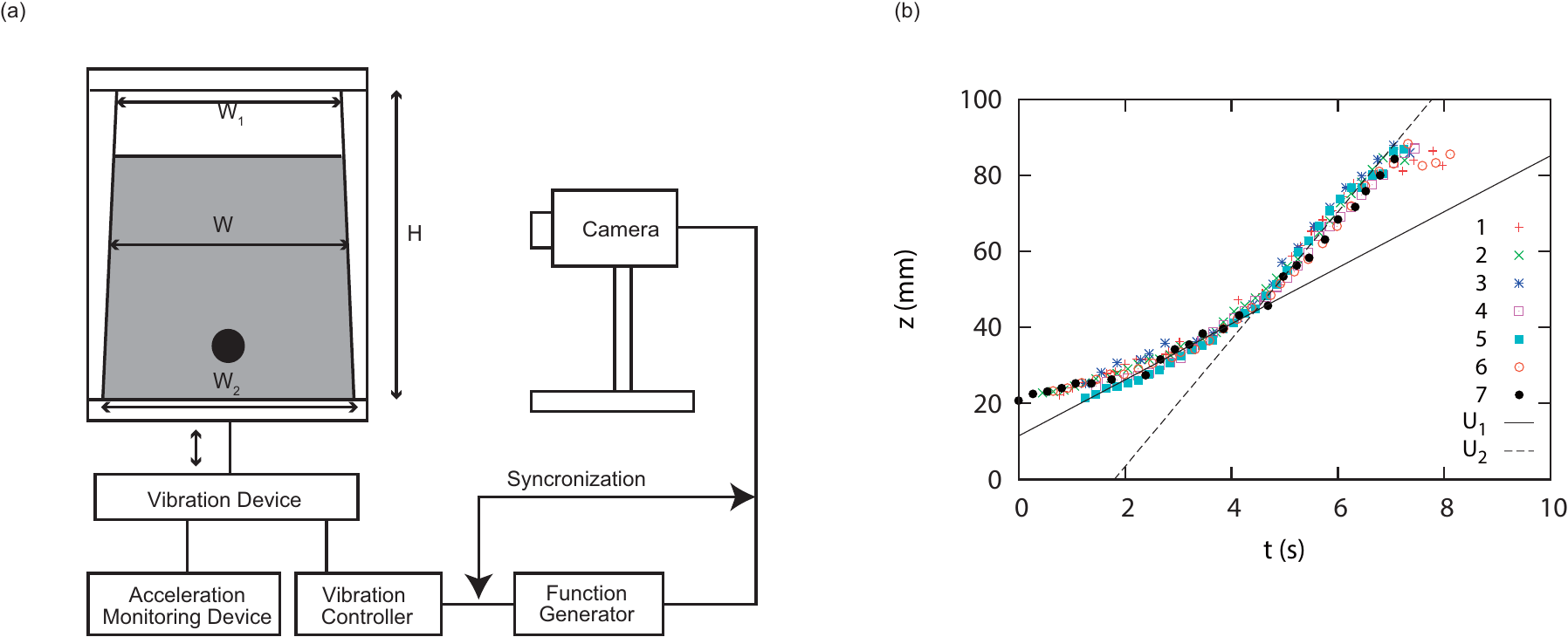}\caption{(a) Experimental setup.
The bottom of the vertically-positioned cell is set as the bottom of the
vertical axis $z$. (b) The position $z$ of the intruder as a function of time
$t$ for the cell with $W=85$ mm and the intruder with $D=12$ mm at the
frequency $f=30$ Hz and the vibration acceleration $a=50$ m/s$^{2}$. The
ascent of the intruder proceeds with two stages. The smaller and larger slopes
indicated by the lines respectively characterize the corresponding two
velocities $U_{1}$ and $U_{2}.$ See the text for the details.}%
\label{Fig1b}%
\end{figure}

The setup is shown in Fig. \ref{Fig1b}a. A cell of thickness 1.2 mm was filled
with beads of average diameter 1.0 mm and a stainless-steel disk of thickness
1.0 mm. The cell is mounted on a vibration system, which causes rising motion
of the disk in granular medium (consisting of one layer of beads) with the
disk sliding freely on the side in the medium.

We used a cell with side walls slightly down-facing with an angle about 5
degrees (specified below) throughout this study. Slight side-wall angle was
necessary to observe stable rising motion of the disk. When the side walls
were vertical, the rising motion was not smooth and not well-reproducible.
Dependence on the angle may originate from a weak friction on the side walls
and a relatively weak polydispersity of our beads (the bead diameter were
approximately in the range 0.9-1.15 mm), which tends to create lattice
structures. With a slight down-facing angle, lattice structures tend to be
broken and the movement of beads near the side walls are restrained as in the
case when friction on the side wall is strong. Note that the importance of
friction on the vertical side wall for convection was pointed out
\cite{1KnightJaegerNagel1993}. They also demonstrated that the introduction an
angle in the opposite up-facing direction can reverse the direction of
convection. However, in the present case, the angle was smaller and introduced
to conduct reproducible study of the BN effect. In this sense, the slight
side-wall angle could be considered as a substitute for polydispersity and
friction near the side walls.

A cell of thickness $t=1.2$ mm was made from two transparent acrylic plates of
thickness 3 mm, separated with spacers of thickness $t$. The cell height $H$
was fixed to 140 mm. The top and bottom widths of the cell, $W_{1}$ and
$W_{2}$ (see Fig. \ref{Fig1b}a), were slightly different with a finite angle
$\alpha$ defined by $\tan\alpha=(W_{2}-W_{1})/(2H)$. The width of the cell $W$
was defined as $W=(W_{1}+W_{2})/2$ for convenience. The cell width $W$ was
either 55, 80, 85, or 105 mm. In the case of $W=85$ mm, $W_{1}$ and $W_{2}$
were set to $W_{1}=80$ and $W_{2}=90$ mm, while $W_{1}$ and $W_{2}$ for
different $W$ were determined such that the angle $\alpha$ was the same with
that for $W=85$ mm, i.e., $\alpha$ was fixed to a value $\alpha=\arctan
[(90-80)/(2\cdot140)]$ ($\sim5$ degrees).

The cell contained a disk-shaped obstacle (diameter $D=12,15$, or 18 mm and
thickness 1 mm) and one layer of small balls of aluminum oxide with an average
diameter $d=1$ mm (AL-9, AS ONE Corp.). The balls filled the cell with the
obstacle to a depth $h=100$ mm. The density of disks ($7.7-7.9$ g/cm$^{3}$)
are larger than that of small alumina balls (3.95 g/cm$^{3}$).

The cell containing the obstacle and grains was mounted vertically on a
vibration generator system (m060/MA1-CE, IMV Corp.), which was controlled by a
multi-function generator (WF1948, NF Corp.) and monitored by an acceleration
meter (VM-1970, IMV Corp.). Digital images were obtained with a CCD camera
(STC-MB33USB, SENTECH Co., Ltd.), which was synchronized with the vibration
system through the function generator.

The vibration generation system can produce continuous sinusoidal waves with a
control on the angular frequency $\omega$ and amplitude $A$. The sinusoidal
wave can be characterized by the vibration velocity $v=\omega A$, acceleration
$a=\omega^{2}A$, and frequency $f=\omega/(2\pi)$.

\section{Results}

\subsection{Rising motion}

As shown in Fig. \ref{Fig1b}b, the intruder rises in the layer of grains with
time under vibration. The different symbols correspond to different ascent
experiments performed under the same condition specified in the caption. As
seen in the plot, all the data points collapse onto a master curve except near
the starting point, demonstrating a reasonable reproducibility of the
experiment. The master curve can be divided into two regimes, characterized by
two velocities $U_{1}$ and $U_{2}$, as indicated in the plot. The linearity in
the second region was more visible than in the first. Determination of the
slopes corresponding $U_{1}$ and $U_{2}$ is explained below in detail. The
crossover depth was around 40 to 60 mm in our parameter ranges, but it was not
sensitive to parameter changes (it was difficult to find a systematic trend).
The initial position was set to $z=20$ mm. When the intruder was initially
placed at a deeper position, the rising motion of the disk became less
reproducible near the bottom. A critical initial depth for rising was
difficult to define because it seemed to be dependent on uncontrollable
initial configurations of beads.

\subsection{Simple laws for $U_{1}$ and $U_{2}$}

\begin{figure}[ptbh]
\includegraphics[width=\textwidth]{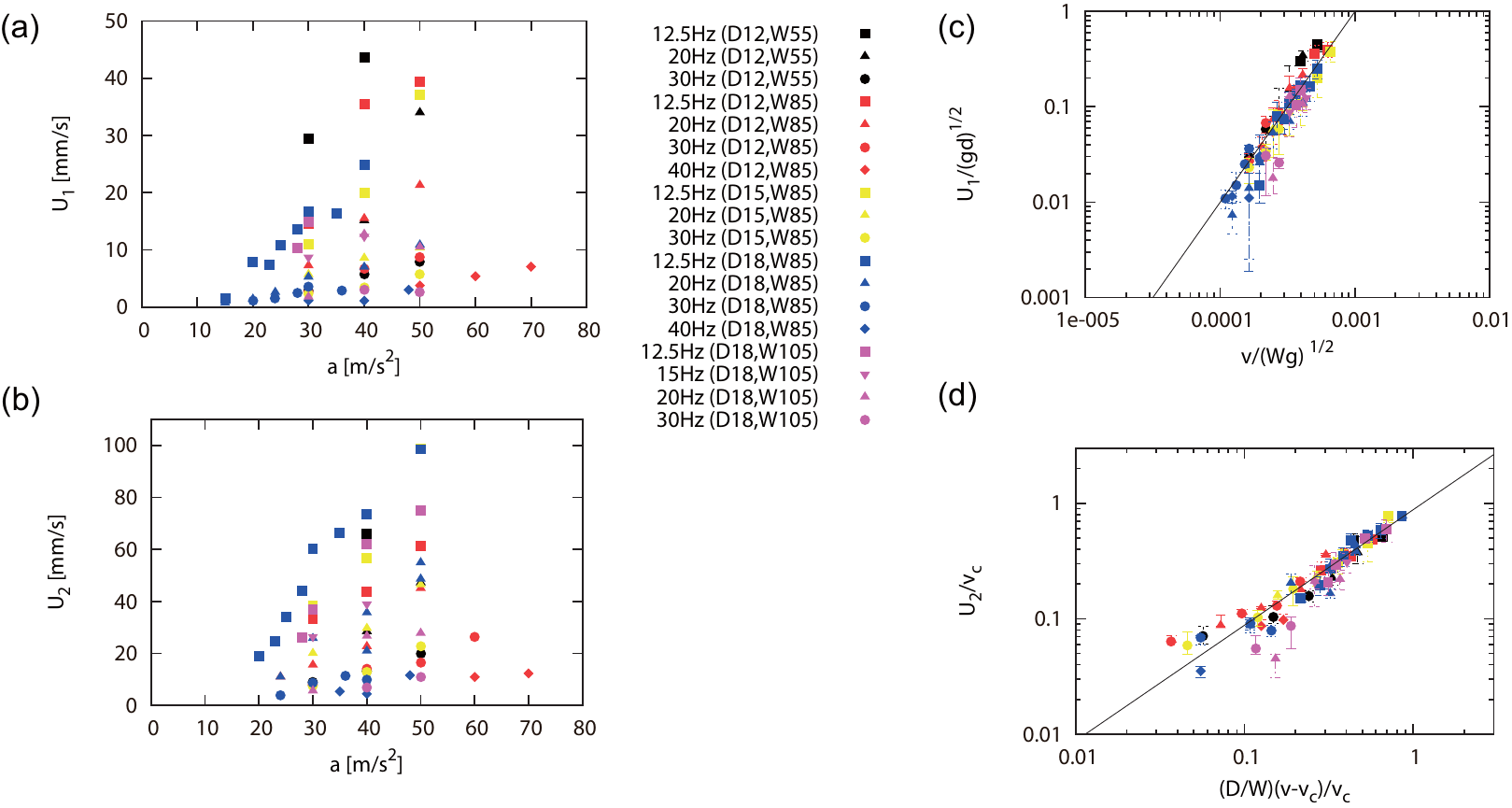}\caption{(a) and (b): The rising
velocity in the first regime $U_{1}$ and that in the second regime as a
function of the vibration acceleration $a$ for different vibration
frequencies, 12.5 Hz (squares), 20 Hz (triangles), 30 Hz (circles) and 40 Hz
(diamonds). Data for $(D,W)=(12,55)$, (12,85), (15,85), (18,85) and (18,105)
are represented by black, red, yellow, blue, magenta symbols, respectively.
(c) All the data in (a) replotted with renormalized axes. The line with a
slope 2 represents Eq. (\ref{eqU1}) with $k_{1}=0.88\pm0.022$. (d) All the
data in (b) replotted with renormalized axes. The line with a slope 1
represents Eq. (\ref{eqU2}) with $k_{2}=0.85\pm0.016$. }%
\label{Fig2}%
\end{figure}

Figure \ref{Fig2} a and b show the two velocities $U_{1}$ and $U_{2}$ obtained
under various conditions as a function of the vibration acceleration $a$. As
demonstrated in Fig. \ref{Fig2} c and d, after carefully looking dependence of
the data on experimental parameters, we found that the data can be well
described by the relations%
\begin{align}
\frac{U_{1}}{\sqrt{gd}}  &  =k_{1}\frac{v^{2}}{Wg}\label{eqU1}\\
\frac{U_{2}}{v_{c}}  &  =k_{2}\frac{D(v-v_{c})}{Wv_{c}} \label{eqU2}%
\end{align}
where $g$ is the gravitational acceleration with $v_{c}=126\pm5$ mm/s,
$k_{1}=0.88\pm0.022$ and $k_{2}=0.85\pm0.016$. The fitting parameter $v_{c}$
was introduced because in the $U_{1}$ vs $v$ plot the data were well on a
straight line and the line intersects with the $v$ axis at the same point
(within experimental errors), irrespective of $D$ and $W$. Note $v_{c}$ is of
the same order of $\sqrt{gd}$, which is a natural velocity scale in the
present problem, as appears in Eq. (1).

The following information could be gleaned from the values of $k_{1}$ and
$k_{2}$. (I) The fact that $k_{1}$ and $k_{2}$ are both close to one confirms
that the order of magnitudes predicted by Eqs. (1) and (2) are consistent with
our experimental data. (II) The fact that the standard deviations for $k_{1}$
and $k_{2}$ thus obtained are less than a few per cent (they are 2.51 and 1.85
\%, respectively) suggests that our experimental data and their analysis are
of high quality.

Details of obtaining values and error bars for Fig. 2 c and d are explained as
follows. We prepared $n$ sets of data under the same experimental condition,
i.e., for a set of $(D,W,f,v)$. For each set of data, we first determined the
second region exploiting its clear linearity with a slope (corresponding to
$U_{2}$), and the data in the remaining region was fit by a straight line with
another slope (corresponding to $U_{1}$). This second fit was conducted by
selecting a linear region as wide as possible with the criterion that the
coefficient of determination became below 0.98 (this coefficient was
determined by $1-\sum_{i=1}^{m}(z_{i}-z_{P})^{2}/\sum_{i=1}^{m}(z_{i}%
-z_{A})^{2}$ where $z_{P}$ and $z_{A}$ are the predicted and average values,
respectively, for $m$ data of $z_{i}$). This well-defined method to determine
$U_{1}$ was used based on the following observations. (1) Rising motion was
unstable near the starting point with less reproducibility. (2) However,
before the second region, which is clearly linear, another linear region
tended to appear and its slope was well reproducible although its width was
not. These features are visible in Fig. 1 b if compares the slopes and the
experimental data. Note that $k_{1}$ and $k_{2}$ used in Fig. 1b are the same
values obtained from Fig. 2c and d. In this way, we obtained $n$ different
sets of ($U_{1}$, $U_{2}$) for a single set of $(D,W,f,v)$. If $n$ was larger
than 10 we employed the standard deviation for the error bar. Otherwise, the
highest and lowest positions of the error bar were determined by the maximum
and minimum values, respectively.

\section{Physical interpretations}

\subsection{First regime: ascent by filling of an arch-shaped void created}

\begin{figure}[ptbh]
\includegraphics[width=\textwidth]{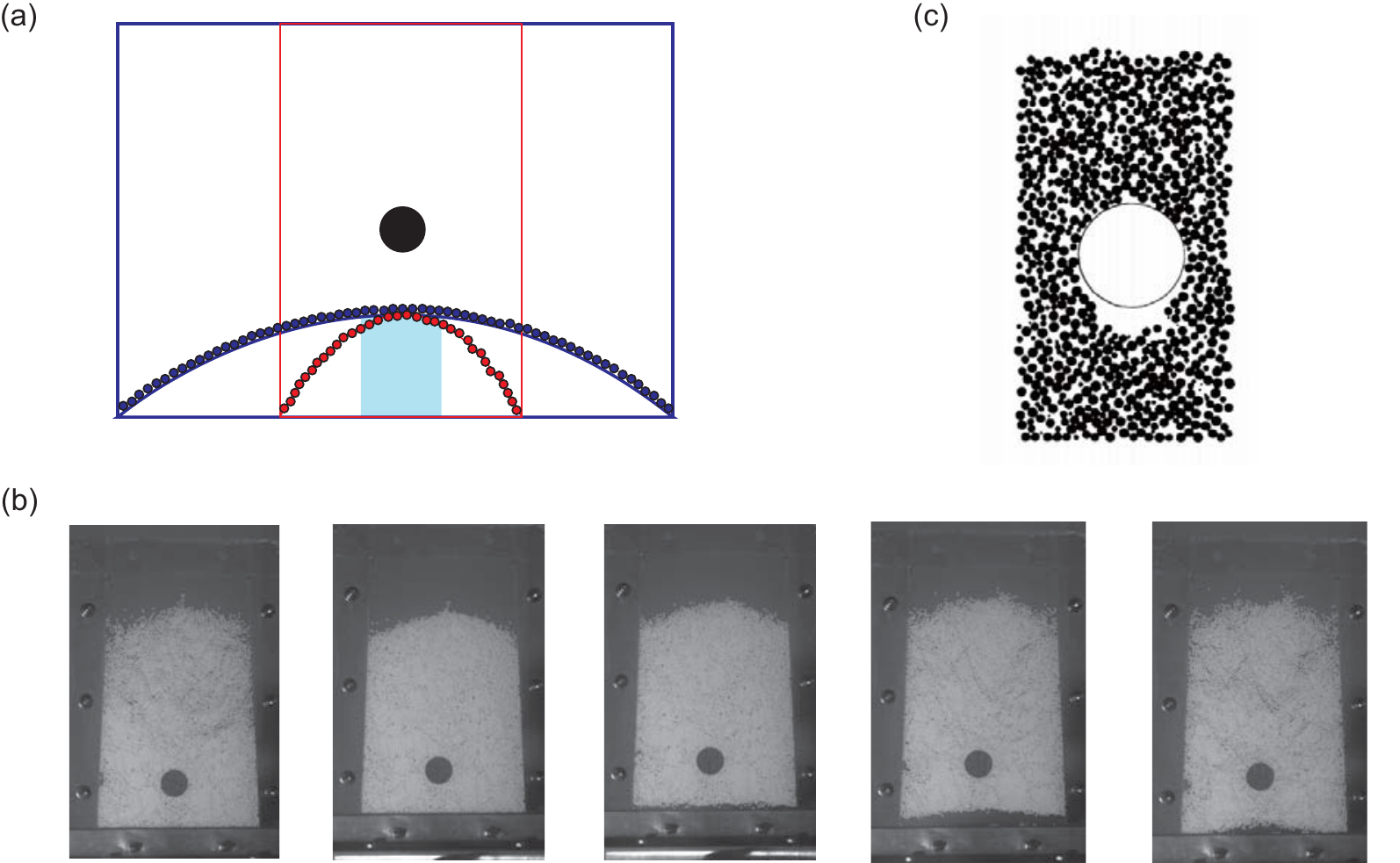}\caption{(a) Illustration of two
arches repeatedly formed at the bottom of the layer during vibration, one in a
cell with a wider cell and another with narrower cell. The width of the arch
is determined by the side walls of the cell. This illustration superimposes
two cases, a wider cell and a narrower cell at the same vibration velocity
when the arch becomes most significant (such a moment repeatedly appear with a
given frequency). The height at the center is the same for the two cells
because it is set by $v^{2}/g$ as explained in the text. (b) Sequential
snapshots, separated by 10 ms, of a cell of $W=85$ mm under vibration at
$f=12.5$ Hz and $a=30$ mm/s$^{2}$ with an intruder of $D=15$ mm. The
arch-shaped void is created at the bottom of the granular layer, which is most
visible in the second right-most photograph. (c) Void filling mechanism
suggested in previous studies: black particles continuously fill up the void
created beneath the intruder during shaking, which results in the rising of
the intruder. (c) is reproduced from Ref. \cite{7jullien1992three} (Copyright,
APS 1992).}%
\label{Fig3}%
\end{figure}

The physical mechanism for rising in the first regime can be considered as a
result of filling by small beads of an arch-shaped void created at the bottom
of the cell during vibration as illustrated in Fig. \ref{Fig3} a. These arch
shapes are created as a result of time-sequential motion as shown in Fig.
\ref{Fig3} b.

For simplicity, we consider the case $a\gg g$ to gain physical insight into
this sequential motion (In fact, the ascent can be observed only when $a>g$,
as also reported in \cite{2duran1994size,4hejmady2012scaling}). When a
sinusoidal wave is applied to the system, the normal force acting on the
system of one grain layer and the intruder (total mass $M$) from the cell
bottom is given by $N=-Mg-M\alpha$, as long as the system holds contact with
the cell bottom, if we neglect the friction effect near the side walls. Here,
$\alpha=-a\cos\omega t$ is the acceleration corresponding to the movement of
the position of the base plate of the cell described by $z=A\cos\omega t$.
This means, when $a\gg g$, soon after the cell moves upwards from the central
position of vibration$,z=0$, the normal $N$ becomes zero and the system starts
a parabolic motion under gravity with the initial speed comparable to
$v=\omega A$. The maximum height of the parabolic motion scales as $h$ with
$v^{2}\simeq gh$ (this relation is exact in Newtonian mechanics, in the
absence of side walls with the constraint that all the particles are on the
vertical plane, and thus tends to hold in the central region if the wall
distance is large and friction with front and back walls is small, which is
the present case). When $a\gg g$, the maximum height $h$ is considerably
larger than the vibration amplitude $A$, and in such a case the maximum height
of the layer bottom relative to the cell bottom is comparable to $h$, which is
the same under a fixed $v$.

However, small beads close to the side walls cannot move (relative to the
cell) because of the friction effect near the side walls, while small beads
around the center of the cell repeatedly go up with the intruder under
vibration till the maximum height comparable to $h\simeq v^{2}/g$ (relative to
the cell bottom) as estimated as above. This leads to the formation of a
velocity gradient in the granular layer in the direction of cell width, which
results in an arch-shaped void at the bottom of the layer as shown by the
snapshots in Fig. \ref{Fig3} b. During the formation of an arch-shaped void,
some beads close to the bottom of the layer enter into the void due to the
velocity gradient, which results in a shear force. As a result, the height of
the intruder relative to the cell bottom when it is next pushed to the cell
bottom becomes higher than before. This is the mechanism of ascent by void filling.

This mechanism of the ascent by void filling is consistent with the dependence
of the ascent velocity $U_{1}$ on the cell width $W$ and on the vibration
velocity $v$, predicted in Eq. (\ref{eqU1}). As long as $a\gg g$, for a fixed
$v$, the maximum height of the arch comparable to $h$ given above is the same,
meaning that the velocity gradient increases as $W$ decreases. Accordingly, we
expect that $U_{1}$ increases as $W$ decreases, which qualitatively explains
why $U_{1}$ scales with $1/W_{1}$ in Eq. (\ref{eqU1}). For a fixed $W$,
because $h$ scales with $v^{2}$, if the void were completely filled, $U_{1}$
would scale with $\omega v^{2}$ by noting that the period of vibration scales
as $1/\omega$. However, the void filling is much less effective, as observed
in Fig. \ref{Fig3} b. Since the effectiveness increases with the period
$\simeq1/\omega$, we expect the ascent per vibration may scale not with
$\omega v^{2}$ but this factor multiplied with the effectiveness factor
$1/\omega$, which results in $U_{1}$ scaling with $v^{2}$, in accordance with
Eq. (\ref{eqU1}). Note that the number of beads falling off from the arch will
increase as the life time of the transient arch increases, i.e., the period of
vibration increases (as shear tends to increase the number). In this sense,
the effectiveness of void filling increases with $1/\omega$ (as $W$ decreases
the effectiveness reducing shear).

According to Eq. (\ref{eqU1}), $U_{1}$ is independent of the disk diameter
$D$. This is in contrast with the ascent ascribed to void filling mechanism in
the previous studies
\cite{5williams1976segregation,6rosato1987brazil,7jullien1992three}. The void
considered in the previous studies is not the arch-shaped void but a void
formed beneath the disk as illustrated in Fig. \ref{Fig3} b. This implies the
ascent velocity increases with $D$. However, in the present case, the size of
the arch-shaped void is independent of the size of the disk (this is because
the width is set by $W$ and the height is set by $v^{2}/g$ as explained
above), and, thus, it is natural that $U_{1}$ is independent of $D$.

\subsection{Second regime: ascent by convection}

\begin{figure}[ptbh]
\includegraphics[width=\textwidth]{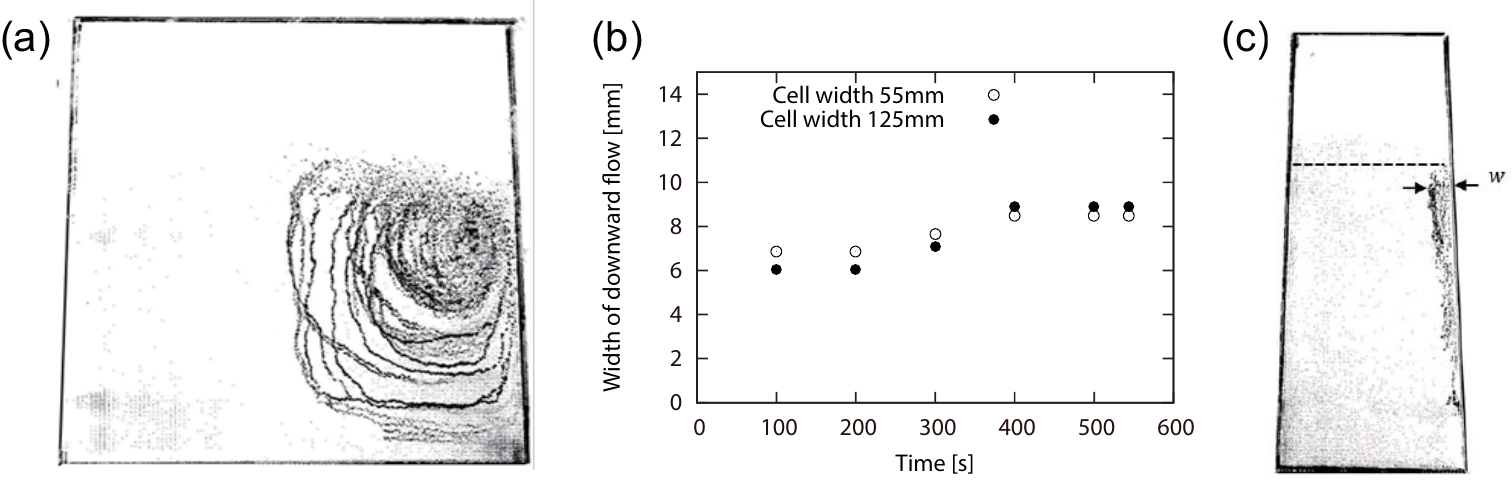}\caption{(a) Convection roll
visualized by superposition of snapshots obtained under the existence of a
single tracer particle in the cell. Images were obtained at the frequency
$f=12.5$ Hz and the acceleration $a=50$ m/s$^{2}$ with a cell with $W=85$ mm.
(b) Width of downward flow developed near a side wall as a function of time,
obtained from two cells with different widths $W=55$ and 125 mm at the same
$f$ and $a$. (c) Superposition of snapshots in first 544 seconds for the cell
with width 55 mm, visualizing the width of downward flow near the side wall.
The dashed horizontal line indicates an average top surface of particles.}%
\label{Fig4}%
\end{figure}

The physical mechanism for rising in the second regime can be considered as a
result of convection. To draw this conclusion, we put a steel bead (silver
color) of the same diameter (1 mm) in a layer of alumina beads (white color)
and track the movement of the steal bead (density 7.85 g/cm$^{3}$). As seen in
a typical result, shown in Fig. \ref{Fig4} a, a convective roll motion is
observed in the right side of the cell. By analyzing the chronological order
of black dots corresponding to the steal bead, it is revealed that the flow is
upward in the center of the cell and is downward near the wall. From symmetry,
a similar roll should exist in the left side of the cell. The upward flow near
the center starts at a certain height where two flows merge, one from the
right side and the other from the left side, and this height seems to
correspond to the height of the transition from the first regime characterized
by $U_{1}$ to the second by $U_{2}$. Accordingly, we can expect that the
rising motion in the second regime is due to the convective roll motion.

For later discussion, we here confirm experimentally that the width of
downward flow developed near a side wall is comparable to the size of grain,
which is consistent with the previous report \cite{1KnightJaegerNagel1993}.
This is shown in Fig. \ref{Fig4} b quantifying the width of the downward flow
near the wall. To obtain this plot, we took snapshots of a vibrated cell with
a single tracer particle and superposed only a selected set of snapshots for a
given duration, where we selected\ only the snapshots in which the particle
moved down compared with its previous snapshot. An example of such
superimposed snapshots is given in Fig. \ref{Fig4} c. We estimated the width
from the maximum width as indicated in the figure. The width thus obtained are
given as a function of time for two cells with different width in Fig.
\ref{Fig4} b. As shown in the plot, the width saturates with time and the
saturated value, which is several times of the grain size, is independent of
the cell width.

\begin{figure}[ptbh]
\includegraphics[width=\textwidth]{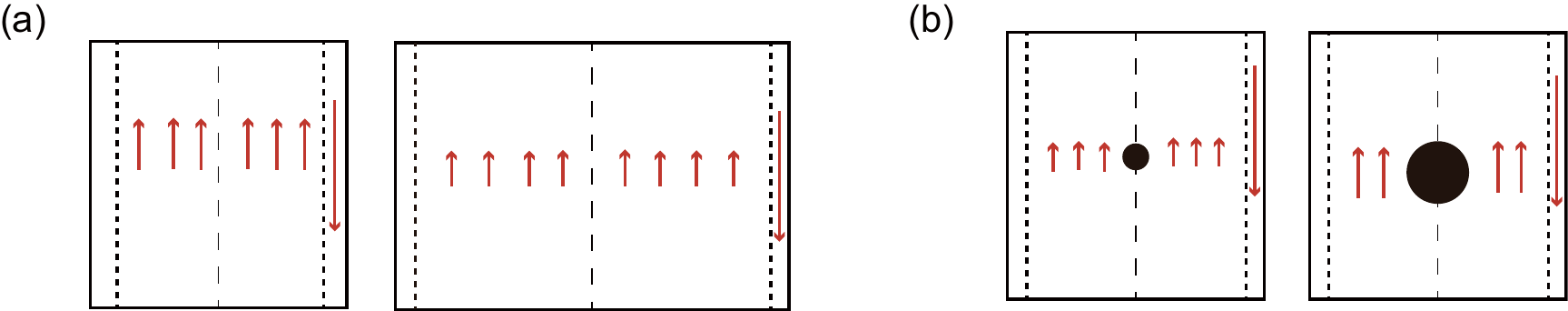}\caption{(a) The effect of cell
width on upward flow velocity. (e) The effect of intruder's diameter on upward
flow velocity. }%
\label{Fig5}%
\end{figure}

The independence of the width of the downward flow near the side wall from the
cell width, confirmed in Fig. \ref{Fig4} b is physically natural because such
a downward flow is caused by the friction effect near the side walls, and this
effect should be the same for cells with different width as long as the side
walls are well separated compared with the width of the downward flow. In
addition, this implies that the flux of the downward flow is independent from
the cell width $W$ for a given $v$, and that the shear force which governs the
flow is characterized not by $v/W$ but by $v/d$.

Considering that the flow flux is conserved in the rolling motion, we expect
that upwards velocity in the absence of the disk (and away from the disk)
scales with a dimensionless factor $d/W$ as illustrated in Fig. \ref{Fig5} a.
However, alongside of the disk, it scales with $d/(W-D)$ as suggested in Fig.
\ref{Fig5} b. In other words, there are two factors $d/W$ and $d/(W-D),$ and
both govern the rising velocity of the disk. Considering that the former is a
decreasing function of $W$ and the latter introduces a factor that increases
with $D$, and expecting that the expression is independent of the smallest
scale, the bead diameter $d$, a simplest possibility incorporating the two
factors is a dimensionless factor $D/W$. This explains why $U_{2}$ scale with
$D/W$ in Eq. (\ref{eqU2}).

This $D$ dependence of $U_{2}$ is in contrast with results reported in the
previous studies \cite{1KnightJaegerNagel1993,4hejmady2012scaling}: they
reported that the velocity of ascent due to convection is independent of the
size of the intruder. This may because the previous cases were more
insensitive to the effect: they both considered not a two dimensional case as
ours but a three-dimensional case, and they considered the cases in which the
ratio of the intruder diameter to grain diameter are larger (the ratio is
$\simeq3.4-306$ in \cite{1KnightJaegerNagel1993} and $\simeq8-13$ in
\cite{4hejmady2012scaling} while it is $\simeq5-7$ in our case).

\section{Conclusion}

We examined the rising motion of an intruder in an original one-layer granular
bed under continuous vibration for a wide range of parameters. We found that
the motion was divided into two regimes and both regimes were well
characterized by vibration velocity ($v$) through simple scaling laws. We
provided physical interpretations based on convection and arch effect.
Although two mechanisms are closely related, they can be clearly
distinguished. The arch effect explains the scaling for the first regime
($\sim v^{2}$), while the convection mechanism explains the second scaling
($\sim v$). These two cases emphasize the importance of vibration velocity for
understanding the BN effect. This generic feature of the BN effect confirmed
in the present study will be useful not only in granular physics but also in
many domain such as agriculture, and cosmetic or pharmaceutical industries
\cite{Duran1997,andreotti2013granular}.

\begin{acknowledgments}
K. O. appreciates Mika Umehara (Ochanomizu University) for giving useful comments. This work was supported by JSPS KAKENHI Grant Number JP19H01859.
\end{acknowledgments}

%\bibliographystyle{naturemagMy}
%\bibliography{C:/Users/okumura/Documents/main/JabRef/granular,C:/Users/okumura/Documents/main/JabRef/fracture,C:/Users/okumura/Documents/main/JabRef/wetting}

\begin{thebibliography}{10}
\expandafter\ifx\csname url\endcsname\relax
  \def\url#1{\texttt{#1}}\fi
\expandafter\ifx\csname urlprefix\endcsname\relax\def\urlprefix{URL }\fi
\providecommand{\bibinfo}[2]{#2}
\providecommand{\eprint}[2][]{\url{#2}}

\bibitem{harwood1977powder}
\bibinfo{author}{Harwood, C.~F.}
\newblock \bibinfo{title}{Powder segregation due to vibration}.
\newblock \emph{\bibinfo{journal}{Powder Technology}}
  \textbf{\bibinfo{volume}{16}}, \bibinfo{pages}{51--57}
  (\bibinfo{year}{1977}).

\bibitem{5williams1976segregation}
\bibinfo{author}{Williams, J.~C.}
\newblock \bibinfo{title}{The segregation of particulate materials. a review}.
\newblock \emph{\bibinfo{journal}{Powder technology}}
  \textbf{\bibinfo{volume}{15}}, \bibinfo{pages}{245--251}
  (\bibinfo{year}{1976}).

\bibitem{6rosato1987brazil}
\bibinfo{author}{Rosato, A.}, \bibinfo{author}{Strandburg, K.~J.},
  \bibinfo{author}{Prinz, F.} \& \bibinfo{author}{Swendsen, R.~H.}
\newblock \bibinfo{title}{Why the brazil nuts are on top: Size segregation of
  particulate matter by shaking}.
\newblock \emph{\bibinfo{journal}{Phys. Rev. Lett.}}
  \textbf{\bibinfo{volume}{58}}, \bibinfo{pages}{1038--1040}
  (\bibinfo{year}{1987}).

\bibitem{clement1992experimental}
\bibinfo{author}{Cl{\'e}ment, E.}, \bibinfo{author}{Duran, J.} \&
  \bibinfo{author}{Rajchenbach, J.}
\newblock \bibinfo{title}{Experimental study of heaping in a two-dimensional
  eesand pileff}.
\newblock \emph{\bibinfo{journal}{Physical Review Letters}}
  \textbf{\bibinfo{volume}{69}}, \bibinfo{pages}{1189} (\bibinfo{year}{1992}).

\bibitem{1KnightJaegerNagel1993}
\bibinfo{author}{Knight, J.~B.}, \bibinfo{author}{Jaeger, H.~M.} \&
  \bibinfo{author}{Nagel, S.~R.}
\newblock \bibinfo{title}{Vibration-induced size separation in granular media:
  The convection connection}.
\newblock \emph{\bibinfo{journal}{Phys. Rev. Lett.}}
  \textbf{\bibinfo{volume}{70}}, \bibinfo{pages}{3728--}
  (\bibinfo{year}{1993}).

\bibitem{duran1993arching}
\bibinfo{author}{Duran, J.}, \bibinfo{author}{Rajchenbach, J.} \&
  \bibinfo{author}{Clement, E.}
\newblock \bibinfo{title}{Arching effect model for particle size segregation}.
\newblock \emph{\bibinfo{journal}{Phys. Rev. Lett.}}
  \textbf{\bibinfo{volume}{70}}, \bibinfo{pages}{2431--2434}
  (\bibinfo{year}{1993}).

\bibitem{7jullien1992three}
\bibinfo{author}{Jullien, R.}, \bibinfo{author}{Meakin, P.} \&
  \bibinfo{author}{Pavlovitch, A.}
\newblock \bibinfo{title}{Three-dimensional model for particle-size segregation
  by shaking}.
\newblock \emph{\bibinfo{journal}{Phys. Rev. Lett.}}
  \textbf{\bibinfo{volume}{69}}, \bibinfo{pages}{640} (\bibinfo{year}{1992}).

\bibitem{knight1996experimental}
\bibinfo{author}{Knight, J.~B.} \emph{et~al.}
\newblock \bibinfo{title}{Experimental study of granular convection}.
\newblock \emph{\bibinfo{journal}{Physical Review E}}
  \textbf{\bibinfo{volume}{54}}, \bibinfo{pages}{5726} (\bibinfo{year}{1996}).

\bibitem{vanel1997rise}
\bibinfo{author}{Vanel, L.}, \bibinfo{author}{Rosato, A.~D.} \&
  \bibinfo{author}{Dave, R.~N.}
\newblock \bibinfo{title}{Rise-time regimes of a large sphere in vibrated bulk
  solids}.
\newblock \emph{\bibinfo{journal}{Physical review letters}}
  \textbf{\bibinfo{volume}{78}}, \bibinfo{pages}{1255} (\bibinfo{year}{1997}).

\bibitem{4hejmady2012scaling}
\bibinfo{author}{Hejmady, P.}, \bibinfo{author}{Bandyopadhyay, R.},
  \bibinfo{author}{Sabhapandit, S.} \& \bibinfo{author}{Dhar, A.}
\newblock \bibinfo{title}{Scaling behavior in the convection-driven brazil nut
  effect}.
\newblock \emph{\bibinfo{journal}{Phys. Rev. E}} \textbf{\bibinfo{volume}{86}},
  \bibinfo{pages}{050301(R)} (\bibinfo{year}{2012}).

\bibitem{yamada2014scaling}
\bibinfo{author}{Yamada, T.~M.} \& \bibinfo{author}{Katsuragi, H.}
\newblock \bibinfo{title}{Scaling of convective velocity in a vertically
  vibrated granular bed}.
\newblock \emph{\bibinfo{journal}{Planetary and Space Science}}
  \textbf{\bibinfo{volume}{100}}, \bibinfo{pages}{79--86}
  (\bibinfo{year}{2014}).

\bibitem{umehara2020rising}
\bibinfo{author}{Umehara, M.} \& \bibinfo{author}{Okumura, K.}
\newblock \bibinfo{title}{Rising obstacle in a two-dimensional granular bed
  induced by continuous and discontinuous vibrations: Dynamics governed by
  vibration velocity}.
\newblock \emph{\bibinfo{journal}{Journal of the Physical Society of Japan}}
  \textbf{\bibinfo{volume}{89}}, \bibinfo{pages}{035001}
  (\bibinfo{year}{2020}).

\bibitem{ehrichs1995granular}
\bibinfo{author}{Ehrichs, E.} \emph{et~al.}
\newblock \bibinfo{title}{Granular convection observed by magnetic resonance
  imaging}.
\newblock \emph{\bibinfo{journal}{Science}} \textbf{\bibinfo{volume}{267}},
  \bibinfo{pages}{1632--1634} (\bibinfo{year}{1995}).

\bibitem{2duran1994size}
\bibinfo{author}{Duran, J.}, \bibinfo{author}{Mazozi, T.},
  \bibinfo{author}{Cl{\'e}ment, E.} \& \bibinfo{author}{Rajchenbach, J.}
\newblock \bibinfo{title}{Size segregation in a two-dimensional sandpile:
  Convection and arching effects}.
\newblock \emph{\bibinfo{journal}{Physical Review E}}
  \textbf{\bibinfo{volume}{50}}, \bibinfo{pages}{5138} (\bibinfo{year}{1994}).

\bibitem{Duran1997}
\bibinfo{author}{Duran, J.}
\newblock \emph{\bibinfo{title}{Sables Poudres et Grains}}
  (\bibinfo{publisher}{Editions Eyrolles in Paris}, \bibinfo{year}{1997}).

\bibitem{andreotti2013granular}
\bibinfo{author}{Andreotti, B.}, \bibinfo{author}{Forterre, Y.} \&
  \bibinfo{author}{Pouliquen, O.}
\newblock \emph{\bibinfo{title}{Granular media: between fluid and solid}}
  (\bibinfo{publisher}{Cambridge University Press}, \bibinfo{year}{2013}).

\end{thebibliography}

\end{document}